\documentclass[fleqn,twoside,twocolumn,nofootinbib]{revtex4} % Specifies the document class %,unsortedaddress
\usepackage[sec]{ujp} % \usepackage[]{ujp} for cyrillic, \usepackage[web]{ujp} for web
\begin{document}
\title[PHENOMENOLOGICAL THEORY OF BOUNDARY FRICTION]%колонтитул
{PHENOMENOLOGICAL THEORY OF BOUNDARY FRICTION IN THE STICK-SLIP MODE}%
\author{I.A.~Lyashenko}%1 автор
\affiliation{Sumy State University}%институт
\address{2, Rimskii-Korsakov Str., Sumy 40007, Ukraine}%адрес
\email{nabla04@ukr.net, khom@mss.sumdu.edu.ua}%e-mail
\author{A.V.~Khomenko}%1 автор
\affiliation{Sumy State University}%институт
\address{2, Rimskii-Korsakov Str., Sumy 40007, Ukraine}%адрес
\email{nabla04@ukr.net, khom@mss.sumdu.edu.ua}%e-mail
\author{L.S.~Metlov}%1 автор
\affiliation{A.A.~Galkin Donetsk Institute for Physics and
Engineering,\\ Nat.
Acad. of Sci. of Ukraine}%институт
\address{72, R. Luxemburg Str., Donetsk 83114, Ukraine}%адрес
\email{lsmet@fti.dn.ua}%e-mail

\udk{539.2:621.891} \pacs{05.70.Ce; 05.70.Ln;\\[-3pt] 47.15.gm;
62.20.Qp; 64.60.-i;\\[-3pt] 68.35.Af; 68.60.-p} \razd{\secix}
\setcounter{page}{278}%
\maketitle

%\makeatletter
%\renewcommand{\thesection}{\arabic{section}}
%\renewcommand{\p@subsection}{}
%\renewcommand{\thesubsection}{\arabic{section}.\arabic{subsection}}
%\renewcommand{\p@subsubsection}{}
%\renewcommand{\thesubsubsection}
%{\arabic{section}.\arabic{subsection}.\arabic{subsubsection}}
%\makeatother

\begin{abstract}
A deterministic theory describing the melting of an ultrathin
lubricant film between two atomically smooth solid surfaces has been
developed. The lubricant state is described by introducing a
parameter of excess volume that arises owing to the
solid structure chaotization at its melting. The thermodynamic and shear
kinds of melting are described consistently. The dependences of
the stationary friction force on the lubricant temperature and the shear
velocity of rubbing surfaces that move with respect to each other
with a constant velocity have been analyzed. In the framework of a
simple tribological model, the stick-slip mode of friction, when the
lubricant periodically melts and solidifies, has been described. The
influence of velocity, temperature, and load on the stick-slip
friction has been analyzed. A qualitative comparison between the
results obtained and experimental data has been carried out.
\end{abstract}

\section{Introduction}

If the thickness of a lubricant material does not exceed 10~atomic
layers, there emerges a boundary friction mode \cite{Persson}.
Experiments show that such thin lubricant layer demonstrates
abnormal properties in comparison with those in the case of bulk
liquids or if the layer thickness is only a few times larger
\cite{Yosh}. In particular, the intermittent (stick-slip) motion
inherent to dry friction is observed \cite{SRC,Yosh}. Such a mode is
explained as a solidification induced by squeezing by rubbing
surfaces, followed by an abrupt melting when shear stresses exceed
the yield stress (\textquotedblleft shear
melting\textquotedblright).

There are a few phenomenological models that partially explain the results
experimentally observed. These include, e.g., thermodynamic
\cite{Popov,1_order}, mechanistic \cite{Carlson}, and synergetic
\cite{KhYu,CondMat} models. They are of either a deterministic
\cite{KhYu,Carlson,1_order} or a stochastic
\cite{Filippov,Filippov1,trenie_iznos} origin. Methods of molecular dynamics
are also applied \cite{Braun,prd,Khome2010}. It turns out that the lubricant
can operate in a number of kinetic modes, and transitions between them take
place in the course of friction, which is responsible for the intermittent
motion \cite{Yosh}. Theoretical researches \cite{Filippov} gave rise to three
modes of friction: a slip mode at low shear velocities, a regular stick-slip
mode, and a slip mode at high shear velocities. Numerous experiments confirm
the existence of those modes \cite{Persson,Yosh,SRC,silica}.

In works \cite{KhYu,CondMat}, in the framework of the Lorentz model
approximating a viscoelastic medium, an approach was developed,
according to which the transition of an ultrathin lubricant film
from a solid-like into a liquid-like state occurs owing to the
thermodynamic and shear melting. The processes invoked by the
self-organization of shear stress and strain fields, as well as the
lubricant temperature, were described with regard for the additive
noises of those quantities \cite{JTPh,dissipative} and correlated
temperature fluctuations \cite{uhl_fnl}. It was shown that, in the
case of additive noises, a self-similar mode of lubricant melting,
in which the series expansion of the stress in time acquires
multifractal properties, is established \cite{3_new,UFJ}. The
reasons for the jump-like melting and hysteresis phenomena, which
were observed in experiments \cite{exp1, exp_n, Isr_rev}, were
examined in works \cite{FTT,PhysLettA,JPS}, and the conditions for
the indicated features to be realized were also determined with
regard for the deformation defect of the shear modulus. In the
framework of this model, the periodic stick-slip friction mode was
also described \cite{period,trenie_iznos}; however, it has a
stochastic component and, hence, can be realized only provided that
there are fluctuations in the system. Another shortcoming of this
model is the fact that it does not involve the load applied to the
friction surfaces; in addition, a number of approximations were
made, while deriving the basic equations \cite{KhYu}.

A thermodynamic theory proposed
in work \cite{1_order} is based on the expansion of the free energy of the system into a
power series in the parameter $f$ which is an excess volume
\cite{Anael1,Anael2} that emerges owing to the formation of a defect
structure in the lubricant at its melting. The liquid-like state is
interpreted as a section of plastic flow on the loading diagram and
is characterized by the presence of defects in the lubricant
\cite{Popov}. In work \cite{1_order}, to describe strongly
nonequilibrium processes that take place at the slipping of two solids
that rub against each other, being separated by a lubricant layer,
an approach based on the Landau theory of phase transitions
\cite{AkadMet,Ran,MFNT,MFNT_Metlov,Metlov_PRE} was used. However,
work \cite{1_order} was devoted to the study of the lubricant melting,
when friction surfaces are moved relative to each other at a
constant velocity, so it does not describe the stick-slip regime of
motion observed in experiments \cite{Yosh}. The proposed work
continues the consideration started in work \cite{1_order}. It aims
at studying the periodic mode of stick-slip friction in the
framework of the model using a mechanical equivalent for the
tribological system, because numerous experimental results evidence
just the periodic character of stick-slip motion~\cite{Yosh}.

\section{Basic Equations}

If the thickness of a melting lubricant is less than 10~molecular
layers, the stationary states, in which the lubricant operates, are
not thermodynamic phases, but represent kinetic modes of friction --
maybe, several ones. In this case, the terms solid-like and
liquid-like phases are used rather that solid and liquid ones.
Whether such lubricants melt or not is judged by the increase of
their volume \cite{Braun} and the diffusion coefficient
\cite{Braun,prd,liqtosol,thompson}. Since, it is the volume that is
an experimentally observable quantity, let us introduce a parameter
$f$ to describe the lubricant state. The physical meaning of this
parameter is an excess volume \cite{Anael1,Anael2} that appears
owing to the stochastization of a solid structure in the course of
its melting. As the parameter $f$ grows, the defect concentration in
the lubricant increases, and the lubricant, governed by the
transport of those defects under the action of an applied tangential
stress, goes into the kinetic mode of plastic flow (the liquid-like
phase).

Let us write down the expansion for the free energy density in the
form which includes the contributions made by
the elastic components of shear strains $\varepsilon_{ij}^{\rm e}$ and
the entropy $s$ \cite{1_order}:
%1
\[
 \Phi = \Phi_0^\ast +
\frac{1}{2}\lambda\left(\varepsilon_{ii}^{\rm e}\right)^2 +
\mu\left(\varepsilon_{ij}^{\rm e}\right)^2 - \alpha s^2 +
\frac{c}{2}\left({\nabla f}\right)^2 -
\]
\begin{equation}- \varphi_0 f + \frac{1}{2}\varphi_1 f^2 - \frac{1}{3}\varphi_2
f^3 + \frac{1}{4}\varphi_3 f^4, \label{int_energy}
\end{equation}
where $\Phi_{0}^{\ast}$, $\lambda$, $\mu$, $\alpha$, $c$, $\varphi_{0}$,
$\varphi_{1}$, $\varphi_{2}$, and $\varphi_{3}$ are expansion constants. In
turn,
%2
\begin{equation}
\varphi_{0}=\varphi_{0}^{\ast}+\frac{1}{2}\bar{\lambda}\left(
\varepsilon _{ii}^{\rm e}\right)  ^{2}+\bar{\mu}\left(
\varepsilon_{ij}^{\rm e}\right)
^{2}+\alpha_{\varphi}s.\label{varphi_const}%
\end{equation}
Elastic stresses are taken into account to an accuracy of quadratic
contributions, expressing them in terms of the strain tensor invariants
$\varepsilon_{ii}^{\rm e}$, $(\varepsilon_{ij}^{\rm e})^{2}=\varepsilon_{ij}%
^{\rm e}\varepsilon_{ji}^{\rm e}$, where the repeated indices mean
summation. In this
case, the first invariant is the trace of the strain tensor, $\varepsilon_{ii}%
^{\rm e}=\varepsilon_{1}^{\rm e}+\varepsilon_{2}^{\rm
e}+\varepsilon_{3}^{\rm e}$, whereas the second one is determined by
the expression~\cite{Kachanov}
%3
\[
(\varepsilon_{ij}^{\rm e})^2 \equiv (\varepsilon_{ll}^{\rm e})^2 -
2I_2 = (\varepsilon_1^{\rm e}+\varepsilon_2^{\rm
e}+\varepsilon_3^{\rm e})^2-
\]
\begin{equation}
-2(\varepsilon_1^{\rm e}\varepsilon_2^{\rm e}+\varepsilon_1^{\rm
e}\varepsilon_3^{\rm e}+ \varepsilon_2^{\rm e}\varepsilon_3^{\rm e})
= (\varepsilon_1^{\rm e})^2+(\varepsilon_2^{\rm
e})^2+(\varepsilon_3^{\rm e})^2.
\end{equation}
According to expression (\ref{int_energy}), elastic stresses
%4
\begin{equation}
\sigma_{ij}^{\rm e}=\frac{\partial\Phi}{\partial\varepsilon_{ij}^{\rm e}}%
=\lambda\varepsilon_{ii}^{\rm
e}\delta_{ij}+2\mu\varepsilon_{ij}^{\rm e}-\left(
\bar{\lambda}\varepsilon_{ii}^{\rm e}\delta_{ij}+2\bar{\mu}\varepsilon_{ij}%
^{\rm e}\right)  f\label{sigma_ij}%
\end{equation}
arise in the lubricant. Expression (\ref{sigma_ij}) can be written in the
form of the effective Hooke law
%5
\begin{equation}
\sigma_{ij}^{\rm e}=2\mu_{\mathrm{eff}}\varepsilon_{ij}^{\rm
e}+\lambda_{\mathrm{eff}}\varepsilon
_{ii}^{\rm e}\delta_{ij}\label{hooke}%
\end{equation}
with effective elastic parameters\footnote{It is necessary to take
$\mu_{\mathrm{eff}}=0$, if $f>\mu/\bar{\mu}$, and
$\lambda_{\mathrm{eff}}=0$,
if $f>\lambda/\bar{\lambda}$.}%
%6
\begin{equation}
\mu_{\mathrm{eff}}  =\mu-\bar{\mu}f,\label{mu_eff}
\end{equation}
%7
\begin{equation}
\lambda_{\mathrm{eff}}  =\lambda-\bar{\lambda}f,\label{lambda_eff}%
\end{equation}
which decrease at the melting, if the parameter $f$ grows.

It is not hard to demonstrate that the invariants are determined as follows:
%8
\begin{equation}
\varepsilon_{ii}^{\rm e}=\frac{n}{\lambda_{\mathrm{eff}}+\mu_{\mathrm{eff}}},\label{first_inv}%
\end{equation}
%9
\begin{equation}
(\varepsilon_{ij}^{\rm e})^{2}=\frac{1}{2}\left[  \left(  \frac{\tau}{\mu_{\mathrm{eff}}%
}\right)  ^{2}+\left(  \varepsilon_{ii}^{\rm e}\right)  ^{2}\right]
,\label{second_inv}%
\end{equation}
where $n$ and $\tau$ are the normal and tangential (shear),
respectively, stress components which are induced in the lubricant
by the rubbing surfaces.\footnote{The shear stresses $\tau$ are
determined by expression (\ref{hooke}) taken at $i\neq j$, i.e., at
$\delta_{ij}=0$. In the case $\mu_{\mathrm{eff}}=0$, the term
$\tau/\mu_{\mathrm{eff}}$ in Eq.~(\ref{second_inv}) should be
replaced, in accordance with Eq.~(\ref{hooke}), by
$2\varepsilon_{ij}^{\rm e}$.} Expressions (\ref{first_inv}) and
(\ref{second_inv}) reflect relations between tensor components and
their invariants in the linear theory of elasticity \cite{Kachanov}.

Let us write down an evolution equation for the nonequilibrium parameter $f$
as the Landau--Khalatnikov equation,
%10
\begin{equation}
\tau_{f}\dot{f}=-\frac{\partial\Phi}{\partial f},\label{newton}%
\end{equation}
where the relaxation time $\tau_{f}$ is introduced. In an explicit form, the equation
reads
%11
\begin{equation}
\tau_{f}\frac{\partial f}{\partial t}=-c\nabla^{2}f+\varphi_{0}-\varphi
_{1}f+\varphi_{2}f^{2}-\varphi_{3}f^{3}-\frac{n^{2}\left(  \bar{\lambda}%
+\bar{\mu}\right)  }{\left(  \lambda_{\mathrm{eff}}+\mu_{\mathrm{eff}}\right)  ^{2}%
},\label{kin_h}%
\end{equation}
where the presence of the last term is associated with the fact that that
invariants (\ref{first_inv}) and (\ref{second_inv}) depend on the excess
volume $f$. In work \cite{1_order}, this term was neglected. However, taking
it into account is important, because, in this case, we can describe the
influence of an external pressure on the melting.

The lubricant temperature is determined from the free energy of the system,
%12
\begin{equation}
T=-\frac{\partial\Phi}{\partial s}=2\alpha s+\alpha_{\varphi}f. \label{T}%
\end{equation}
Therefore, the entropy is a function of the temperature and the excess volume. In this
case, the free energy (\ref{int_energy}) is also a function of the temperature and the volume.

To describe the processes of heat exchange between the lubricant and its
environment, let us introduce the temperature of friction surfaces, $T_{e}$
\cite{KhYu}. If the medium is heated up non-uniformly, the heat
equation is an ordinary continuity equation \cite{Landau}
%13
\begin{equation}
T\frac{\partial s}{\partial t}=\kappa\nabla^{2}T, \label{teplo}%
\end{equation}
where the heat conductivity coefficient $\kappa$ is adopted to be constant.
For the normal component $\nabla_{z}^{2}$, the approximation $\kappa\nabla
_{z}^{2}T\approx(\kappa/h^{2})(T_{e}-T)$, where $h$ is the lubricant
thickness, can be used with a sufficient accuracy. Then, Eq.~(\ref{teplo})
reads
%14
\begin{equation}
\frac{\partial s}{\partial t}=\frac{\kappa}{h^{2}}\left(  \frac{T_{e}}%
{T}-1\right)  +\frac{\kappa}{T}\nabla_{xy}^{2}T, \label{teplo_new}%
\end{equation}
where the quantity $h^{2}/\kappa$ plays the role of the relaxation time, during
which the temperatures becomes uniform across the lubricant thickness due to
the heat conductivity.

Let us take advantage of the Debye approximation, which couples elastic,
$\varepsilon_{ij}^{\rm e}$, and plastic, $\varepsilon_{ij}^{\rm
pl}$, strains \cite{Popov},
%15
\begin{equation}
\dot{\varepsilon}_{ij}^{\rm pl}=\frac{\varepsilon_{ij}^{\rm
e}}{\tau_{\varepsilon}},
\label{e_pl}%
\end{equation}
where $\tau_{\varepsilon}$ is the Maxwell time of internal stress relaxation.
The total strain in the layer is determined as
%16
\begin{equation}
\varepsilon_{ij}=\varepsilon_{ij}^{\rm e}+\varepsilon_{ij}^{\rm pl}.
\end{equation}
This strain governs the motion velocity of the upper block, $V_{ij}$,
according to the following relation \cite{Wear}:
\begin{equation}
%17
V_{ij}=h\dot{\varepsilon}_{ij}=h(\dot{\varepsilon}_{ij}^{\rm
e}+\dot{\varepsilon
}_{ij}^{\rm pl}). \label{V}%
\end{equation}
Three last relations yield an expression for the elastic component of
a shear strain \cite{1_order},
%18
\begin{equation}
\tau_{\varepsilon}\dot{\varepsilon}_{ij}^{\rm e}=-\varepsilon_{ij}^{\rm e}%
+\frac{V_{ij}\tau_{\varepsilon}}{h}. \label{e_ij_solid}%
\end{equation}
Below, to simplify the analysis, we consider a uniform system, so
that $\nabla\equiv0$ has to be put in relations (\ref{int_energy}),
(\ref{kin_h}), and (\ref{teplo_new}).

\section{Friction Force}

The system of kinetic equations (\ref{kin_h}), (\ref{teplo_new}), and
(\ref{e_ij_solid}), taking definitions (\ref{varphi_const}), (\ref{hooke}%
)--(\ref{second_inv}), and (\ref{T}) into account, is closed and can
be used to study the melting kinetics. In this section,
we consider the stationary modes of friction. According to
Eqs.~(\ref{teplo_new}) and (\ref{e_ij_solid}), the stationary values
of lubricant temperature $T_{0}$ and the elastic component of the shear
strain $\varepsilon_{ij0}^{\rm e}$ are established in due course,
%19
\begin{equation}
T_{0}=T_{e},\quad\varepsilon_{ij0}^{\rm
e}=\frac{V_{ij}\tau_{\varepsilon}}{h}.
\label{st}%
\end{equation}
To find the stationary states of all quantities, it is necessary to
numerically solve the evolution equation (\ref{kin_h}), using formulas
(\ref{varphi_const}), (\ref{hooke})--(\ref{second_inv}), and determining the
current entropy from Eq.~(\ref{T}), taken at $T=T_{e}$, and the strain from
Eq.~(\ref{st}).

In experimental works, the dependences of the friction force on the shear
velocity, thickness of a lubricant layer, and normal pressure are often
reported \cite{Persson,Yosh,SRC,exp1,exp_n,Isr_rev}. In this section, we
analyze the influence of the lubricant temperature and the shear velocity on
the friction force.

In the lubricant, besides elastic stresses, $\sigma_{ij}^{\rm e}$,
the viscous ones, $\sigma_{ij}^{\mathrm{visc}}$, also emerge. The total
stress in the layer is a sum of those two contributions,
%20
\begin{equation}
\sigma_{ij}=\sigma_{ij}^{\rm e}+\sigma_{ij}^{\rm visc}.\label{sigma_all}%
\end{equation}
The full friction force is determined in a standard way,
%21
\begin{equation}
F_{ij}=\sigma_{ij}A,\label{F_begin}%
\end{equation}
where $A$ is the area of contacting surfaces. The viscous stresses in the
layer are expressed by the formula \cite{Wear}
%22
\begin{equation}
\sigma_{ij}^{\rm visc}=\frac{\eta_{\mathrm{eff}}V_{ij}}{h},\label{sigma_ij_v_begin}%
\end{equation}
where $\eta_{\mathrm{eff}}$ is the effective viscosity which can be found
only experimentally and in the boundary regime \cite{Wear,visc_Book},
%23
\begin{equation}
\eta_{\mathrm{eff}}\sim\left(  \dot{\varepsilon}_{ij}\right)  ^{\gamma},\label{eta_eff}%
\end{equation}
where $\gamma<0$ for pseudo-plastic lubricants, and $\gamma>0$ for dilatant
ones \cite{visc_Book}. Taking Eqs.~(\ref{V}) and (\ref{eta_eff}) into account,
expression (\ref{sigma_ij_v_begin}) for viscous stresses reads
%24
\begin{equation}
\sigma_{ij}^{\rm visc}=\left(  \frac{V_{ij}}{h}\right)  ^{\gamma+1}%
.\label{sigma_ij_v}%
\end{equation}
After substituting Eqs.~(\ref{sigma_all}) and (\ref{sigma_ij_v}) into
Eq.~(\ref{F_begin}), we obtain the sought expression for the friction force
\cite{1_order}\footnote{Here, we use both the sign function, $\mathrm{sgn}%
(x)$, and the absolute value of shear velocity, $|V_{ij}|$, because the
velocity can accept negative values as well.}
%25
\begin{equation}
F_{ij}=\left[  \sigma_{ij}^{\rm e}+\mathrm{sgn}(V_{ij})\left(  \frac{|V_{ij}|}%
{h}\right)  ^{\gamma+1}\right]  A,\label{F}%
\end{equation}
where the quantity $\sigma_{ij}^{\rm e}$ is given by formula
(\ref{hooke}) at $i\neq j$.

As experimental friction surfaces, atomically smooth mica surfaces are usually
used, whereas quasispherical molecules of octamethylcyclotetrasiloxane
(OMCTS) and linear chain molecules of either tetradecane or hexadecane play
the role of a lubricant \cite{Yosh,Isr_rev}. The indicated experiments are
carried out under the following conditions: the thickness of a lubricant layer
$h\approx10^{-9}$~m, the contact area $A\approx3\times10^{-9}~\mathrm{m}^{2}$, and
the load on the upper friction surface lies within the interval $L=(2\div
60)\times10^{-3}~\mathrm{N}$, which corresponds to normal stress
$n=-L/A=-(6.67\div200)\times10^{5}$~Pa. The friction force is $F\approx
(2\div40)\times10^{-3}~\mathrm{N}.$ In the course of the cited
experimental works, it was revealed that the lubricant melts, if either the
temperature exceeds the critical value, $T_{e}>T_{c0}\approx300~\mathrm{K}$,
or the shear velocity $V$ is larger than $V_{c}\approx400$\textrm{~nm/s}.
Those thresholds can substantially vary, depending on the lubricant type and
the experimental geometry.

In the model concerned, according to experimental data, the following values
of theoretical constants were selected \cite{1_order}: $\Phi_{0}^{\ast}%
=20$~J/m$^{3}$, $\lambda=2\times10^{11}$~Pa,
$\bar{\lambda}=10^{8}$~Pa, $\mu=4.1\times10^{11}$~Pa,
$\bar{\mu}=4\times10^{11}$~Pa, $\varphi_{0}^{\ast }=5$~J/m$^{3}$,
$\varphi_{1}=1100$~J/m$^{3}$, $\varphi_{2}=2700$~J/m$^{3}$,
$\varphi_{3}=2070$~J/m$^{3}$,
$\alpha=0.055$~K$^{2}{\cdot}$m$^{3}$/J, $\alpha_{\varphi}=0.05$~K,
$h=10^{-9}$~m, $\tau_{f}=1$~$\mathrm{Pa}{\cdot }\mathrm{s}$, and
$\tau_{\varepsilon}=10^{-8}$~s. Note that the time of the excess
volume relaxation $\tau_{f}$ has the dimension of viscosity.
Actually, this means that the time of the establishment of a
stationary friction mode increases with the effective viscosity of a
a lubricant.

Dependence (\ref{F}) is shown in Fig.~1. Figure 1,$a$ illustrates
the fact that the friction force decreases, as the temperature
grows. Let us consider curve \textit{2} in more details. At first,
the excess volume monotonously grows together with the temperature.
On the contrary, the effective shear modulus $2\mu_{\mathrm{eff}}$
(see Eq.~(\ref{mu_eff})) decreases, which results in a decrease of
the elastic component of shear stresses (\ref{hooke}) and,
accordingly, in a reduction of the friction force (\ref{F}). When
the temperature exceeds the critical value, $T_{e}>T_{c0}$, the
magnitude of excess volume $f$ increases in a jump-like manner, and
the lubricant melts, which causes a drastic decrease of the total
friction force. If the lubricant temperature decreases further, the
lubricant solidifies already at a lower value $T_{e}=T_{c}^{0}$. The
temperature dependence has a hysteretic character, which corresponds
to the first-order phase transitions. According to Fig.~1,$a$, if
the shear velocity increases, the lubricant melts at a lower
temperature. At a velocity higher than a certain critical value, the
lubricant is always liquid-like, irrespective of the temperature
(curve \textit{4}), and the friction force decreases together with
the temperature owing to a reduction of the shear modulus (lubricant
fluidization).

%Fig. 1
\begin{figure}% figure* for wide figure, [h] [!] to change the placement
\includegraphics[width=6.5cm]{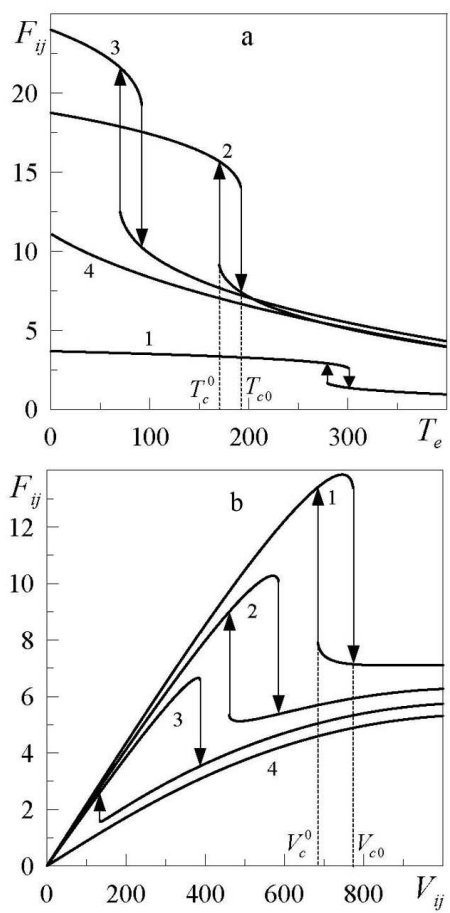}
\vskip-3mm\caption{Dependences of the stationary value of total
friction force {$F_{ij}$} [mN] (see formula {(\ref{F})}) on the
friction surface temperature {$T_{e}$} [K] ($a$) and on the shear
velocity {$V_{ij}$} [nm/s] ($b$) at {$\gamma=2/3$,
$A=3\times10^{-9}$~m$^{2}$, and $n=-7\times10^{5}$~Pa}. Curves 1 to
4 in panel $a$ correspond to the constant shear velocity
${V_{ij}=150}$, 800, 1100, and 1400~nm/s, respectively; curves 1 to
4 in panel $b$ correspond to the fixed temperatures ${T_{e}=200}$,
245, 279, and 310$~\mathrm{K}$, respectively  }
\end{figure}

Hence, at low temperatures ($T_{e}<T_{c}^{0}$), the potential
$\Phi(f)$ has one minimum which corresponds to a stationary state
with small $f$ (a solid-like lubricant). In the temperature interval
$T_{c}^{0}<T_{e}<T_{c0}$, two minima of $\Phi(f)$ coexist. However,
the system cannot go into the state that corresponds to the second
minimum, because those minima are separated by a potential barrier.
If the temperature increases further, $T_{e}>T_{c0}$, the separating
maximum disappears, and the lubricant, following the mechanism of
the first-order phase transition, goes into a state that corresponds
to a unique minimum at large $f$, i.e., it melts. If then the
temperature is reduced, the appearance of the first minimum cannot
force the system to transit back into the corresponding state
because of the separating maximum. When the latter disappears at
$T_{e}=T_{c}^{0}$, the lubricant undergoes a jump-like
solidification.

Figure 1,$b$ illustrates a somewhat different behavior. Here, in
accordance with Eq.~(\ref{F}), the lubricant is solid-like, if the
shear velocities are low, and the corresponding $\sigma_{ij}^{\rm
e}$-values are large. The velocity growth in such a mode gives rise
to an increase of both friction force components (\ref{F});
therefore, the friction force increases rapidly. If the velocity
grows further, the lubricant melts, and the elastic shear stress
(\ref{hooke}) decreases considerably, which results, in turn, in a
drastic reduction of the total friction force. If the velocity
increases even more, the value of $F_{ij}$ grows again, driven by
both friction force components which increase together with the
shear velocity. In the liquid-like state, according to curve
\textit{4}, the friction force (\ref{F}) also grows owing to an
increase in the velocity. That is, we have a situation which is
similar to the behavior of the system, when the temperature is
elevated (Fig.~1,$a$); the difference is that, in the case of the
shear melting, the growth of the friction surface temperature is
accompanied by an increase of the area confined by the hysteresis
loop. As the temperature increases, the lubricant melts at lower
shear velocities. We emphasize that the results depicted in
Fig.~1,$b$ qualitatively coincide with a new friction map for the
boundary lubrication proposed in work \cite{Wear} as a
generalization of experimental data. No dependences of the friction
force on the temperature -- of the type exhibited in Fig.~1,$a$ --
are measured experimentally now.\looseness=1

\section{Stick-Slip Mode}

The dependences depicted in Fig.~1 were obtained at a fixed shear
velocity of the upper friction surface. However, the dynamic
characteristics of the tribological system are governed not only by
the friction force presented in this figure, but also by the
properties of the system in whole. In particular, according to
experiments, in the hysteresis section of the dependence shown in
Fig.~1, an intermittent (stick-slip) mode of friction is possible
\cite{Yosh,Carlson,Filippov,Filippov1,Braun,Isr_rev,liqtosol,Wear},
and this work is devoted to the elucidation of its features. A
typical scheme of tribological system is shown in Fig.~2. A spring,
the rigidity of which is $k$, is connected with a block of mass $M$,
to which an additional loading $L$ is applied. The block is located
on a smooth surface, being separated from it by a lubricant layer of
the thickness $h$. The free end of the spring is forced to move with
a constant velocity $V_{0}$. When the block moves, there emerges a
friction force $F$ (\ref{F}) that interferes with the block motion.
In the case of ultrathin lubricant layers operating under the
boundary friction conditions, the block, $V$, and spring, $V_{0}$,\
velocities can differ from each other because of the oscillating
character of the force $F$, which results in the stick-slip motion
of the block. This mode is similar to dry friction (with no
lubricant).\looseness=1

%Fig. 2
\begin{figure}% figure* for wide figure, [h] [!] to change the placement
\includegraphics[width=\column]{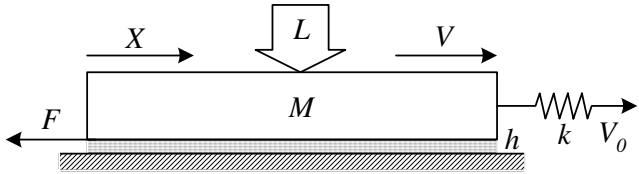}
\vskip-3mm\caption{Scheme of a tribological system  }
\end{figure}

The equation of motion for the upper block looks like
\cite{Yosh,Popov,Carlson}\footnote{Since a unidirectional shear is considered,
tensor notations are omitted hereafter for convenience.}
%26
\begin{equation}
M\ddot{X}=k\left(  \int\limits_{0}^{t}{V_{0}}dt^{\prime}-X\right)
-F,\label{Move}%
\end{equation}
where $t=t^{\prime}$ is the time of motion. If the shear velocity
$V_{0}$ is constant, the integral in expression (\ref{Move}) can be
substituted by $V_{0}t,$ of course. In order to calculate the time
evolution of the friction force, this equation has to be solved
together with Eqs.~(\ref{kin_h}), (\ref{teplo_new}), and
(\ref{e_ij_solid}); then, the friction force is determined by
expression (\ref{F}). However, since the lubricant layer is thin,
the relaxation times of strain, $\tau_{\varepsilon}$, and entropy,
$\tau_{s}=h^{2}/\kappa$, can be considered short in comparison with
the relaxation time of excess volume $\tau_{f}$. Therefore, within
the approximation $\tau_{f}\gg\tau_{\varepsilon},\tau_{s}$, we have
to solve two equations, (\ref{Move}) and (\ref{kin_h}),
consistently; then, the temperature and strain can be determined
from Eq.~(\ref{st}), and the entropy from Eq.~(\ref{T}).

The solution of the specified equations is depicted in Fig.~3. The figure
demonstrates that the friction force monotonously grows at first, because the
lubricant is solid-like, and the shear velocity increases. When the latter
exceeds the critical value $V_{c0}$, the lubricant melts. As a result, the
friction force decreases, the motion velocity of the upper block $V$ grows, and
the block shifts over a large distance. The spring tension and, accordingly,
the shear velocity decrease at that. When the latter becomes less than it is
necessary to keep the lubricant in the liquid-like state, the lubricant
solidifies, and the friction force starts to grow. The described process is
periodically repeated. Note that the velocity, at which the lubricant
solidifies, does not coincide with the analogous velocity presented in Fig.~1.
This fact is associated with a drastic growth of the shear velocity $V$ at the melting
and the corresponding increase of the parameter $f$. According to
Eq.~(\ref{mu_eff}), the shear modulus becomes negative, so it is necessary to
consider it as zero-valued, which changes the form of potential
(\ref{int_energy}). Then, in the presence of elastic strains (\ref{e_ij_solid}%
), the elastic stresses in the lubricant, according to formula (\ref{hooke}),
vanish, which is responsible for the friction force reduction, so that the
lubricant can flow.

%Fig. 3
\begin{figure}% figure* for wide figure, [h] [!] to change the placement
\includegraphics[width=7cm]{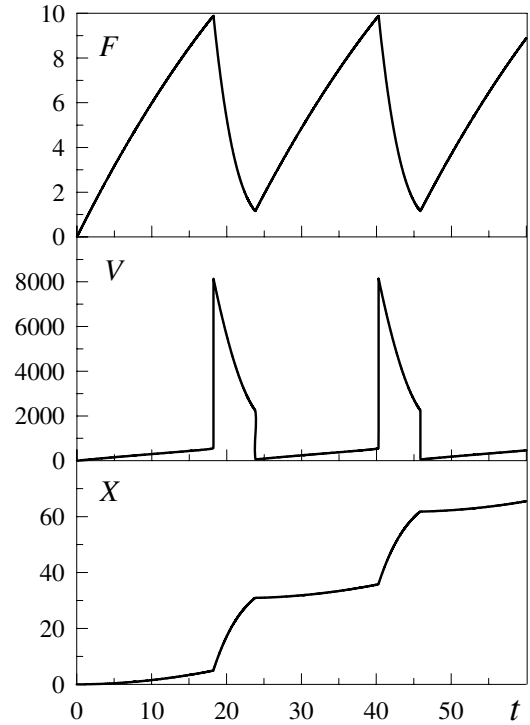}
\vskip-3mm\caption{Dependences of the friction force {$F$ }[mN]
($a$), shear velocity $V$ of the friction surface [nm/s] ($b$), and
coordinate $X$ of the friction surface [$\mu\mathrm{m}$] on the time
$t$ [s] at {$n=-7\times10^{5}$~Pa}, $M=0.4$~kg, $k=480$~N/m,
$T_{e}=250~\mathrm{K}$, and $V=1400$\textrm{~nm/s}  }
\end{figure}

%Fig. 4
\begin{figure}% figure* for wide figure, [h] [!] to change the placement
\includegraphics[width=7cm]{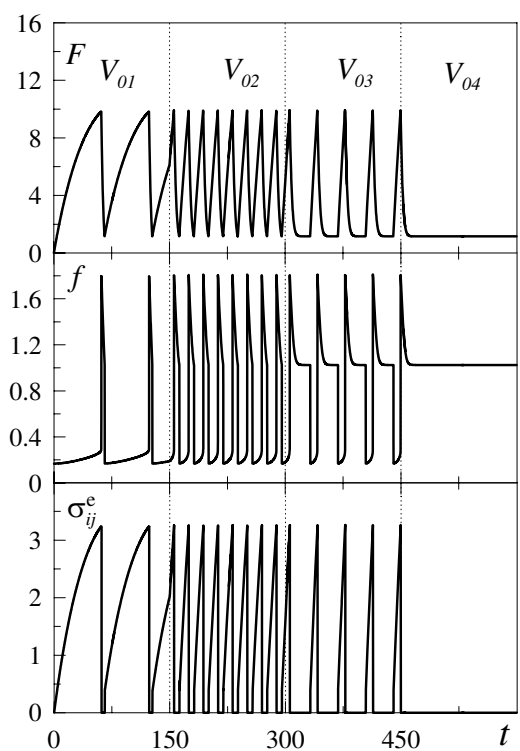}
\vskip-3mm\caption{Dependences of the friction force $F$ [mN] ($a$),
excess volume $f$ ($b$), and elastic stress component
{$\sigma_{ij}^{\rm e}$} [MPa] ($c$) on the time $t$ [s] for various
shear velocities {$V_{01}=650$~nm/s, $V_{02}=1800$~nm/s,
$V_{03}=2246.7$~nm/s, and $V_{04}=2247$~nm/s. The other }parameters
are the same as in Fig.~3  }
\end{figure}

In Fig.~4, the time dependences of the total friction force $F$
(\ref{F}), excess volume $f$, and elastic component of shear
stresses $\sigma _{ij}^{\rm e}$ (\ref{hooke}) at
various--increasing--values of shear velocity
$V_{0}$ are shown. First, the motion of the upper movable block ($V_{0}%
=V_{01}$) results in the growth of the excess volume $f$. When, the
$f$-value reaches the critical magnitude, the lubricant melts
following the mechanism of the first-order phase transitions, so
that the parameter $f$ increases stepwise. Afterwards, the lubricant
starts to solidify again, because the relative shear velocity of
friction surfaces decreases (see Fig.~3). After the lubricant has
totally solidified, an elastic stress appears in it. The further
growth of the stress leads to a new increase of the parameter $f$,
until the latter reaches the critical value necessary for melting,
and the process repeats again. As a result, a periodic intermittent
(stick-slip) regime melting/solidification is established.

When the velocity is increased to the value $V_{0}=V_{02}$, the frequency of
the stick-slip peak appearance also increases, because the critical $f$-value is
attained more rapidly at this velocity. Respectively, the lubricant melts more
quickly; therefore, the system undergoes a larger number of
melting/solidification transitions within the same time interval.

A further velocity growth to $V_{0}=V_{03}$ results in a reduction of
the stick-slip peak frequency. This occurs, because there appear long kinetic
sections in the dependence $F(t)$, where $F=\mathrm{const}$. It should be
noted that, in this mode, the parameter $f$ drastically grows first at the melting
owing to a rapid increase of the shear velocity $V$ of the upper block. The
stationary kinetic section is associated with a smaller excess volume $f$,
which is established after the sharp initial shift of the upper rubbing surface.
This shift is induced by a partial release of the mechanical potential energy
stored by the stretched spring.

As the shear velocity increases to $V_{0}=V_{04}$, the
stick-slip mode disappears, and a kinetic friction regime for the
liquid-like lubricant is established, which is characterized by a
larger value of excess volume $f$ and zero elastic stresses
$\sigma_{ij}^{\rm e}$. Note that the liquid-like state is not always
characterized by zero stresses $\sigma_{ij}^{\rm e}$ \cite{1_order}.
In our case, this fact stems from the equality of the effective
shear modulus of a lubricant (\ref{mu_eff}) to zero in the liquid-like
state \cite{Popov}.

%Fig. 5
\begin{figure}% figure* for wide figure, [h] [!] to change the placement
\includegraphics[width=6.8cm]{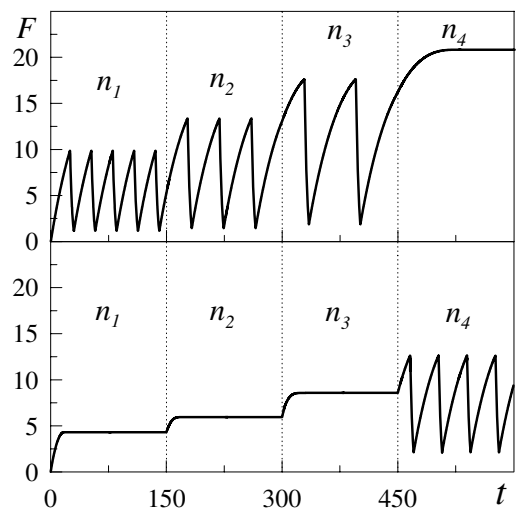}
\vskip-3mm\caption{Dependences of the friction force $F$ [mN] on the
time $t$ [s] at the same parameters as in Fig.~3 and for various
external normal loads {$n_{1}=-7\times10^{5}$~Pa,
$n_{2}=-50\times10^{5}$~Pa, $n_{3}=-80\times
10^{5}$~Pa, and $n_{4}=-100\times10^{5}$~Pa and various }temperatures ${T}%
_{e}=250$ (upper panel) and 400$~\mathrm{K}$ (lower panel)  }
\end{figure}

Hence, when the shear velocity grows, the frequency of stick-slip
peaks increases firstly; then, it decreases owing to the
appearance of long kinetic sections. If the critical velocity
$V_{0}$ is exceeded, the stick-slip mode disappears. Such a
behavior agrees well with experimental data \cite{Yosh}.

In experiments, the influence of an external pressure applied normally to the
friction surfaces on the melting of a lubricant is also often studied
\cite{Yosh,liqtosol}. Such experiments testify that the pressure affects the
parameters of a tribological system in a nontrivial manner. For example, the
critical shear velocity decreases as the pressure grows for lubricants
containing hexadecane chain molecules and, on the contrary, increases for
quasispherical OMCTS ones \cite{Yosh}. The pressure also influences the
frequency and the amplitude of stick-slip transitions \cite{Yosh}. In the
framework of our model, according to Eq.~(\ref{kin_h}), the growth of a loading
on the friction surface results in a reduction of the excess volume, which should
favor the lubricant solidification.

Figure~5 illustrates the time dependences of the friction force at
various values of normal pressure that squeezes the friction surfaces.
At a temperature lower than the critical value (the upper panel in
the figure), the stick-slip mode of friction is realized. In this
case, the pressure growth is accompanied by an increase of the
amplitude of stick-slip transitions and the magnitudes of kinetic
and static friction forces, as well as by a reduction of the
transition frequency. At a pressure that corresponds to the normal
stress $n=n_{4}$, the stick-slip regime does not take place.
However, the kinetic mode, which corresponds to the liquid-like
lubricant, is not established. Instead, the lubricant solidifies
owing to its squeezing by walls. As a result, the lubricant cannot
melt any more, and a large friction force $F$ is established, which
corresponds to the case of a solid-like lubricant and a small value of
excess volume $f$, because the squeezing of the lubricant by walls
promotes the emergence of an atomic long-range order in it.

The lower panel of Fig.~5 exhibits the time dependence of the friction force that
is observed at an elevated temperature of friction surfaces, $T_{e}$. One can
see that the kinetic mode of friction is established, which corresponds to low
values of friction force and large values of excess volume $f$. However, at
$n=n_{4}$, the stick-slip mode starts, because, according to Eq.~(\ref{kin_h}%
), the lubricant cannot always be liquid-like at such values of normal stress
$n$. If the pressure grows further, one has to expect a complete
solidification of a lubricant, as it takes place in the upper panel of the
figure at $n=n_{4}$. Therefore, we revealed three friction modes: 1)~a kinetic
mode, in which the lubricant is always liquid-like, 2)~a stick-slip mode,
which corresponds to periodic melting/solidification cycles, and 3)~a dry
friction mode, which is characterized by a large value of friction force and a
solid-like structure of the lubricant. These modes were also found in work
\cite{Filippov} in the framework of a stochastic model.

In Fig.~6, the time dependences of the friction force at various temperatures of
friction surfaces, which coincide in this consideration with the lubricant
temperatures, are shown. One can see that the temperature elevation results in
a decrease of the friction force oscillation amplitude and in a frequency growth
for the phase transitions ``liquid-like lubricant--solid-like lubricant''. At $T_{e}=T_{e4}%
$, the slip mode starts, which is characterized by a constant value
of the kinetic friction force and a constant shear velocity of the
upper block. Hence, the temperature growth promotes the lubricant
melting. This dependence is a prediction, because we do not know
about the experiments of this type aimed at studying the temperature
effects.

%Fig. 6
\begin{figure}% figure* for wide figure, [h] [!] to change the placement
\includegraphics[width=7cm]{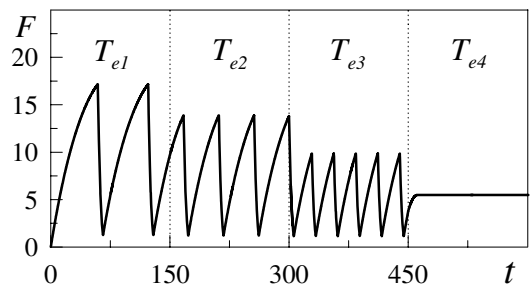}
\vskip-3mm\caption{Dependences of the friction force $F$ [mN] on the
time $t$ [s] at the same parameters as in Fig.~3 and for various
temperatures of friction surfaces: $T_{e1}=150$~K, $T_{e2}=200$~K,
$T_{e3}=250$~K, and $T_{e4}=300$~K  }
\end{figure}

\section{Conclusions}

The proposed theory allows one to describe the effects that are
observed at the melting of an ultrathin lubricant film in the
boundary friction mode. A consistent consideration of the
thermodynamic and shear melting has been carried out. The
dependences of the friction force on the shear velocity and the
temperature have been examined. At high temperatures of friction
surfaces, the shear melting begins at lower values of shear
velocities (shear stresses), and, if the temperature grows
further, the lubricant melts even at the zero shear velocity. In
the model proposed, the influences of the temperature, the shear
melting, and the external pressure are taken into account. Those
parameters are the main factors, which are studied experimentally.

In the framework of the proposed theory, a simple tribological
system is studied, and the time dependences of the friction force
are obtained for increasing the shear velocity, pressure, and
temperature. It is shown that, in a wide range of parameters of
the system, the intermittent friction mode is realized, which is
observed experimentally. The pressure is found to affect the
system in a nontrivial way. The results obtained qualitatively
coincide with known experimental data. Since the model is
quantitative, its modifications can be used for the description of
specific experiments.

\vskip3mm The authors are grateful to the State Fund of Fundamental
Researches of Ukraine and to the Russian Fund of Fundamental
researches (grant Ф28/443-2009) for their sponsorship.

\rezume{%
ФЕНОМЕНОЛОГIЧНА ТЕОРIЯ ПЕРЕРИВЧАСТОГО\\ РЕЖИМУ МЕЖОВОГО ТЕРТЯ}{Я.О.
Ляшенко, О.В. Хоменко, Л.С. Метлов} {Побудовано детерміністичну
теорію плавлення ультратонкої плівки мастила,  яку затиснуто між
двома атомарно-гладкими твердими поверхнями. Для опису стану мастила
введено параметр надлишкового об'єму, що виникає за рахунок
хаотизації структури твердого тіла у процесі плавлення. Узгоджено
описано термодинамічне і зсувне плавлення. Проаналізовано залежності
стаціонарної сили тертя від температури мастила і швидкості зсуву
поверхонь, що труться, при їх рівномірному зсуві зі сталою
швидкістю. У межах простої трибологічної моделі описано
переривчастий режим тертя, при якому мастило періодично плавиться і
твердне. Проаналізовано вплив швидкості, температури і навантаження
на переривчасте тертя. Проведено якісне порівняння отриманих
результатів із експериментальними даними.}

\end{document}